\documentclass[cup6b,epsf]{cupbook}
\usepackage{epsf}
\usepackage{psfig}
\usepackage{natbib}

\def\simlt{\mathrel{\hbox{\rlap{\hbox{\lower4pt\hbox{$\sim$}}}\hbox{$<$}}}}
\def\simgt{\mathrel{\hbox{\rlap{\hbox{\lower4pt\hbox{$\sim$}}}\hbox{$>$}}}}

\begin{document}

\author{Edo Berger \\
Division of Physics, Mathematics and Astronomy, California Institute
of Technology, \\ Pasadena, CA 91125}

\chapter{The Diversity of Cosmic Explosions: $\gamma$-Ray Bursts and
Type Ib/c Supernovae}

{\it The death of massive stars and the processes which govern the
formation of compact remnants are not fully understood.
Observationally, this problem may be addressed by studying different
classes of cosmic explosions and their energy sources.  Here we
discuss recent results on the energetics of $\gamma$-ray bursts (GRBs)
and Type Ib/c Supernovae (SNe Ib/c).  In particular, radio
observations of GRB\,030329, which allow us to undertake calorimetry
of the explosion, reveal that some GRBs are dominated by mildly
relativistic ejecta such that the total explosive yield of GRBs is
nearly constant, while the ultra-relativistic output varies
considerably.  On the other hand, SNe Ib/c exhibit a wide diversity in
the energy contained in fast ejecta, but none of those observed to
date (with the exception of SN\,1998bw) produced relativistic ejecta.
We therefore place a firm limit of $3\%$ on the fraction of SNe Ib/c
that could have given rise to a GRB.  Thus, there appears to be clear
dichotomy between hydrodynamic (SNe) and engine-driven (GRBs)
explosions.}

\section{The Death of Massive Stars}

The death of massive stars ($M\simgt 8$ M$_\odot$) is a chapter of
astronomy that is still being written.  Recent advances in modeling
suggests that a great diversity can be expected.  Indeed, such
diversity has been observed in the neutron star remnants: radio
pulsars, AXPs, and SGRs.  We know relatively little about the
formation of black holes.

The compact objects form following the collapse of the progenitor
core.  The energy of the resulting explosion can be supplemented or
even dominated by the energy released from the compact object (e.g.~a
rapidly rotating magnetar or an accreting black hole).  Such "engines"
can give rise to asymmetrical explosions (MacFadyen \& Woosley 1999),
but even in their absence the core collapse process appears to be
mildly asymmetric (e.g.~Wang et al. 2001). Regardless of the source of
energy, a fraction, $E_K$, is coupled to the debris or ejecta (mass
$M_{\rm ej}$) and it is these two gross parameters which determine the
appearance and evolution of the resulting explosion.  Equivalently one
may consider $E_K$ and the mean initial speed of ejecta, $v_0$, or the
Lorentz factor, $\Gamma_0=[1-\beta_0^2]^{-1/2}$, where $\beta_0=
v_0/c$.

Supernovae (SNe) and $\gamma$-ray bursts (GRBs), are distinguished by
their ejecta velocities.  In the former $v_0\sim 10^4$ km s$^{-1}$ as
inferred from optical absorption features (e.g.~Filippenko 1997),
while for the latter $\Gamma_0\simgt 100$, inferred from the
non-thermal prompt emission (Goodman 1986; Paczynski 1986),
respectively.  The large difference in initial velocity arises from
significantly different ejecta masses: $M_{\rm ej}\sim few$ M$_\odot$
in SNe compared to $\sim 10^{-5}$ M$_\odot$ in GRBs.

In the conventional interpretation, $M_{\rm ej}$ for SNe is large
because $E_K$ is primarily derived from the (essentially) symmetrical
collapse of the core and the energy thus couples to all the mass left
after the formation of the compact object.  GRB models, on the other
hand, appeal to a stellar mass black hole remnant, which accretes
matter on many dynamical timescales and powers relativistic jets (the
so-called collapsar model; MacFadyen \& Woosley 1999).

Still, as demonstrated by the association of the energetic Type Ic
SN\,1998bw ($d\approx 40$ Mpc) with GRB\,980425 (Galama et al. 1998),
as well as the association of SN\,2003dh with GRB\,030329 (e.g.~Stanek
et al. 2003), some overlap may exist.  Here we take an observational
approach to investigating the diversity of stellar explosions, their
energetics, and the relation between them focusing in particular on
GRBs and SNe Ib/c.

\section{The Energetics of $\gamma$-Ray Bursts}

\begin{figure}[htbp]
\centerline{\psfig{file=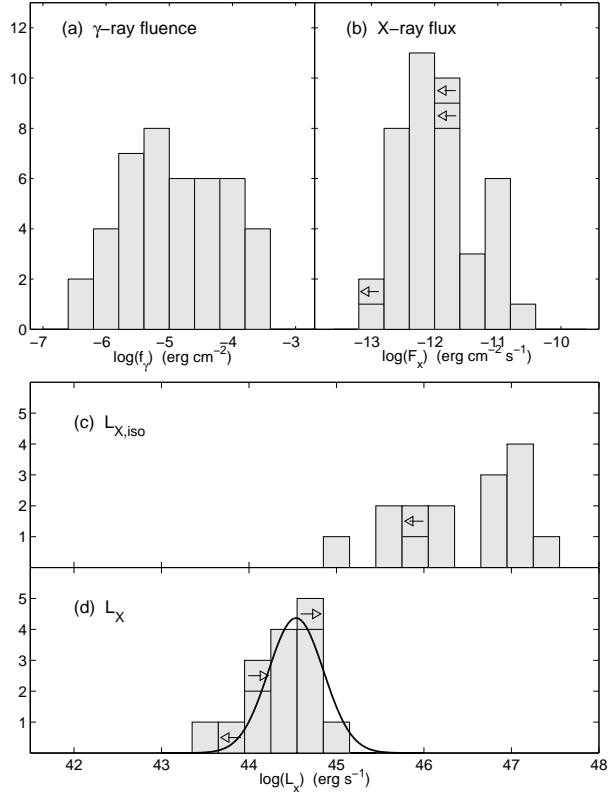,width=8cm}}
\caption[]{\small (a) Distribution of $\gamma$-ray fluences; (b)
Distribution of X-ray fluxes scaled to $t=10$ hr after the burst; (c)
Isotropic-equivalent X-ray luminosity plotted for the subset of X-ray
afterglows with known jet opening angles and redshifts; (d) True X-ray
luminosity corrected for beaming, a proxy for the afterglow kinetic
energy.}
\label{fig:xray}
\end{figure}

Recent studies revealed the surprising result that long-duration GRBs
have a standard energy of $E_\gamma\approx 1.3\times 10^{51}$ erg in
ultra-relativistic ejecta when corrected for asymmetry (``jets'';
Frail et al. 2001; Bloom, Frail \& Kulkarni 2003).  A similar result
was found for the kinetic energies of GRB afterglow using the
beaming-corrected X-ray luminosities as a proxy for the true kinetic
energy (Figure~\ref{fig:xray}; Berger, Kulkarni \& Frail 2003).
However, these studies have also highlighted a small group of
sub-energetic bursts, including the peculiar GRB\,980425 associated
(Galama et al. 1998) with SN\,1998bw ($E_\gamma\approx 10^{48}$ erg).
Until recently, the nature of these sources has remained unclear.

\begin{figure}[htbp]
\centerline{\psfig{file=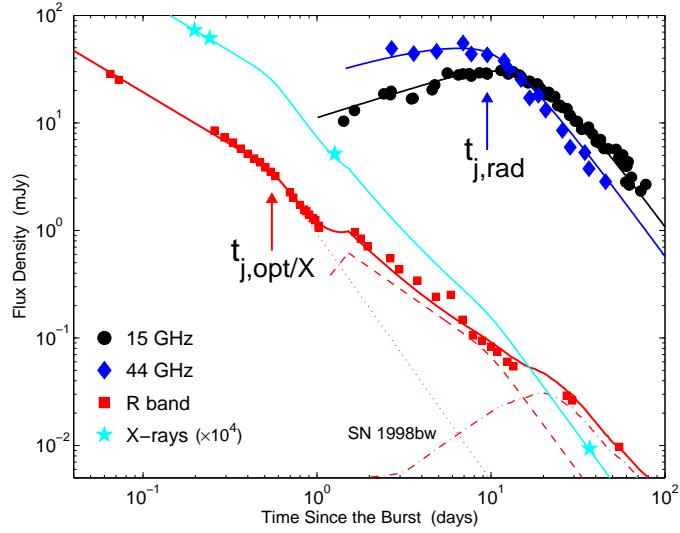,width=9cm}}
\caption[]{\small Radio to X-ray lightcurves of the afterglow of
GRB\,030329, exhibiting the early jet break at 0.55 days in the
optical and X-ray bands, as well as the subsequent rise in optical
flux and the bright radio emission arising from a jet with an opening
angle of $17^\circ$.  Also plotted is the optical emission from
SN\,1998bw at the redshift of GRB\,030329 as a proxy for SN\,2003dh.
The solid line is a combination of all the different emission
components.}
\label{fig:030329}
\end{figure}

This question appears to now be resolved thanks to broad-band
calorimetry of GRB\,030329 (Berger et al. 2003a), the nearest
cosmological burst detected to date (redshift, $z=0.1685$).  Early
optical observations of the afterglow of GRB\,030329 revealed a sharp
break at $t=0.55$ day (Figure~\ref{fig:030329}; Price et al. 2003).
The X-ray flux (Tiengo et al. 2003) tracks the optical afterglow for
the first day, with a break consistent with that seen in the optical.
Thus, the break at 0.55 day is not due to a change in the ambient
density since for typical parameters (e.g.~Kumar 2000) the X-ray
emission is not sensitive to density.  However, unlike the optical
emission the X-ray flux at later time continues to decrease
monotonically.

Given the characteristic $F_\nu\propto t^{-2}$ decay for both the
X-ray and optical emission beyond 0.55 day, the break is reasonably
modeled by a jet with an opening angle of $5^\circ$.  The inferred
beaming-corrected $\gamma$-ray energy is only $E_\gamma\approx 5\times
10^{49}$ erg, significantly lower than the strong clustering around
$1.3\times 10^{51}$ erg seen in most bursts.  Similarly, the
beaming-corrected X-ray luminosity at $t=10$ hours is $L_{X,10}\approx
3\times 10^{43}$ erg s$^{-1}$, a factor of ten below the tightly
clustered values for ``typical'' bursts (Figure~\ref{fig:xray}).

\begin{figure}[htbp]
\centerline{\psfig{file=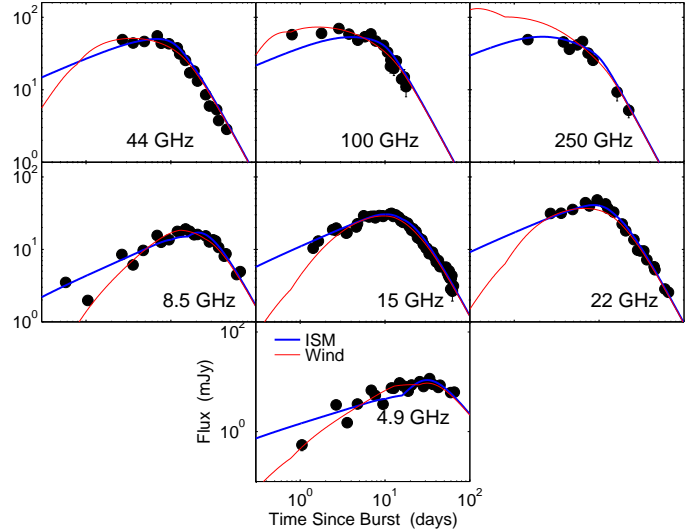,width=9cm}}
\caption[]{\small Detailed radio lightcurves of the afterglow of
GRB\,030329.  The solid lines are afterglow models of collimated
ejecta expanding into circumburst media with a uniform and wind
($\rho\propto r^{-2}$) density profiles.  The sharp turnover at about
10 days is indicative of a jet with an opening angle of $17^\circ$.}
\label{fig:030329r}
\end{figure}

The radio afterglow of GRB\,030329 (Berger et al. 2003a; Sheth et
al. 2003) reveals a different picture.  The increase in flux during
the first 10 days, followed by a rapid decline, $F_\nu\propto
t^{-1.9}$ at $t\simgt 10$ day and a decrease in peak flux at
$\nu\simlt 22.5$ GHz (Figure~\ref{fig:030329r}) are indicative of a
jet with an opening angle of 17$^\circ$.  The inferred
beaming-corrected kinetic energy in the explosion is about $3\times
10^{50}$ erg, comparable to what is inferred from modeling of other
afterglows (Panaitescu \& Kumar 2002).

This result, combined with the resurgence in the optical emission at
1.5 days, is best explained in the context of a two-component
explosion model.  In this scenario the first component (a narrow jet,
$5^\circ$) with initially a larger Lorentz factor is responsible for
the $\gamma$-ray burst and the early optical and X-ray afterglow
including the break at 0.55 day, while the second component (a wider
jet, $17^\circ$) powers the radio afterglow and late optical emission
(Figure~\ref{fig:030329}; Berger et al. 2003a).  The break at 10 days
due to the second component has recently been inferred in the optical
bands following a careful subtraction of the light from SN\,2003dh
which accompanied GRB\,030329 (e.g.~Matheson et al. 2003).  Such a
two-component jet finds a natural explanation in the collapsar model
(Zhang, Woosley \& Heger 2003).

\begin{figure}[htbp]
\centerline{\psfig{file=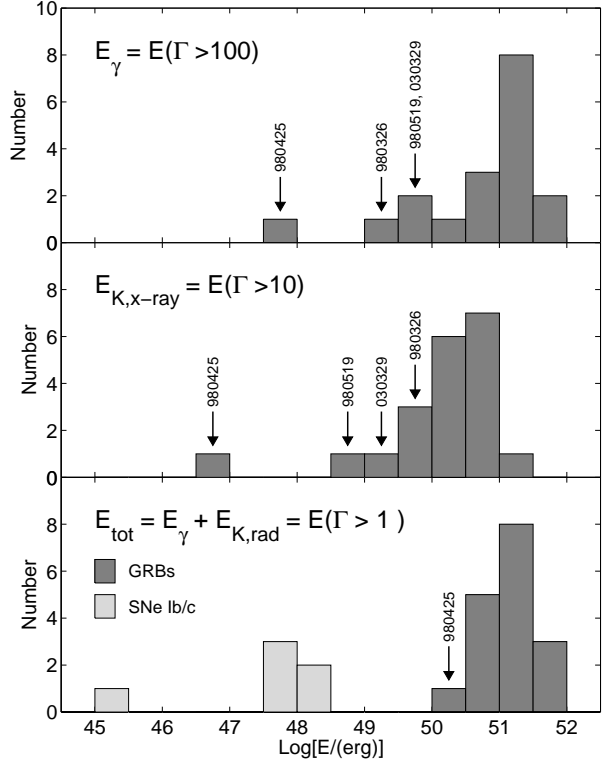,width=8cm}}
\caption[]{\small The beaming-corrected energies of GRBs and SNe Ib/c
in ejecta with a Lorentz factor ranging from $\simgt 100$ to order
unity ({\it top:} the $\gamma$-ray energy, {\it middle:} the kinetic
energy in the early afterglow; {\it bottom:} the total relativistic
energy).  The ultra-relativistic output of GRBs varies considerably
despite a nearly standard total explosive yield.  On the other hand,
the significantly lower total energy in fast ejecta of SNe Ib/c points
to a different energy source.}
\label{fig:energy}
\end{figure}

The afterglow calorimetry of GRB\,030329 has important ramifications
for our understanding of GRB engines and the sub-energetic bursts.
Namely, such bursts may have a total explosive yields similar to other
GRBs (Figure~\ref{fig:energy}), but their ultra-relativistic output
varies considerably.

This leads to the following conclusions. First, radio calorimetry,
which is sensitive to all ejecta with $\Gamma\simgt few$, shows that
the total energy yield of GRB\,030329 is similar to those estimated
for other bursts.  Along these lines, the enigmatic GRB\,980425
associated with the nearby supernova SN\,1998bw also has negligible
$\gamma$-ray emission, $E_{\gamma,{\rm iso}}\approx 8\times 10^{47}$
erg; however, radio calorimetry (Li \& Chevalier 1999) showed that
even this extreme event had a similar explosive energy yield
(Figure~\ref{fig:energy}).  The newly recognized class of cosmic
explosions, the X-ray Flashes, exhibit little or no $\gamma$-ray
emission but appear to have comparable X-ray and radio afterglows to
those of GRBs (see \S\ref{sec:future}).  Thus, the commonality of the
total energy yield points to a common origin, but apparently the
ultra-relativistic output is highly variable.  Unraveling what
physical parameter is responsible for this variation appears to be the
next frontier in the field of cosmic explosions.

\section{The Incidence of Engine in Type Ib/c Supernovae}

The inferences summarized in the previous section, coupled with the
association of some GRBs with SNe Ib/c raises the question: is there a
population of SNe that is powered by engines?  Observationally there
appear to be many distinctions (e.g.~ejecta velocity and mass), but
the association of the Type Ic SN\,1998bw with GRB\,980425 has
indicated that some overlap exists.  In particular, the radio emission
from SN\,1998bw revealed mildly relativistic ejecta with a complex
structure indicative of a long-lived energy source.  The expected
fraction of similar events in the local SN population, $\sim 0.5\%$ to
$25\%$, depends on their origin: typical GRBs observed away from the
jet axis versus an intermediate population of explosions.

\begin{figure}[htbp]
\centerline{\psfig{file=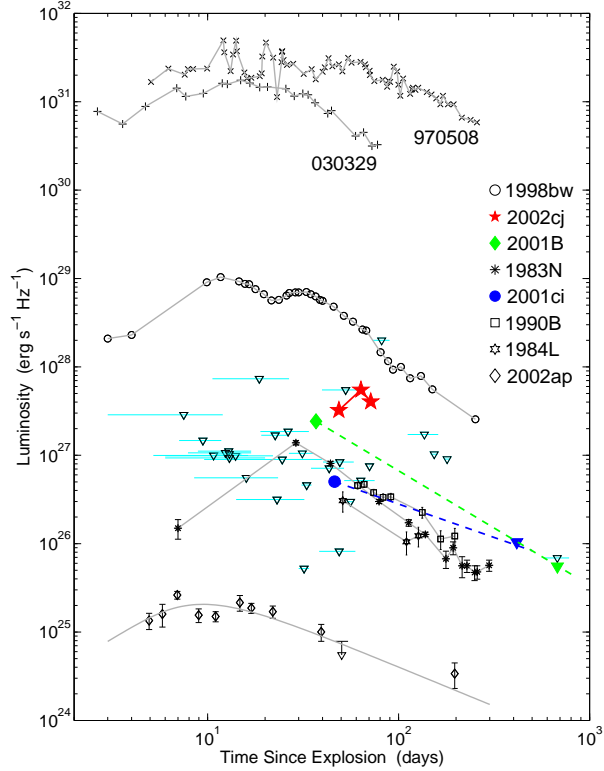,width=8cm}}
\caption[]{\small Radio lightcurves of Type Ib/c SNe including the
peculiar SN\,1998bw/GRB\,980425.  A comparison to SN\,1998bw and GRB
afterglows reveals significantly less energy in high velocity ejecta,
and thus constrains the fraction of SNe Ib/c that are powered by an
engine to $<3\%$.  There is therefore a dichotomy in the explosion
mechanism of massive stars.}
\label{fig:snlcs}
\end{figure}

To assess this fraction, and hence the origin of potential
engine-driven SNe, directly, we have carried out since 1999 the most
comprehensive radio survey of SNe Ib/c to date (Berger, Kulkarni \&
Chevalier 2002; Berger et al. 2003b).  As was demonstrated in the case
of SN\,1998bw, such observations provide the best probe of
relativistic ejecta (a proxy for an engine).

\begin{figure}[htbp]
\centerline{\psfig{file=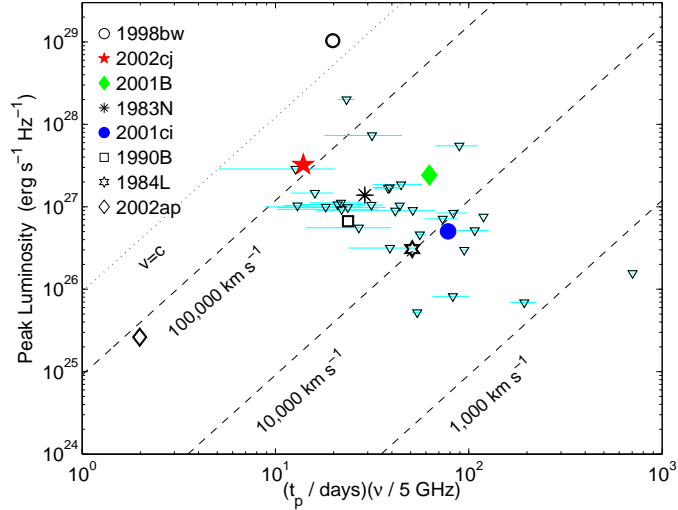,width=9cm}}
\caption[]{\small Peak radio luminosity plotted versus the time of
peak luminosity for Type Ib/c SNe.  The diagonal lines are contours of
constant average expansion velocity based on the assumption that the
peak of the radio luminosity occurs at the synchrotron self-absorption
frequency.}
\label{fig:snvel}
\end{figure}

As seen in Figure~\ref{fig:snlcs} the luminosity function of SNe Ib/c
is significantly broader than previously inferred, but none of the
observed SNe approach the luminosity of typical GRB afterglows.  We
therefore place a limit of about $3\%$ on the fraction of local SNe
Ib/c that are powered by an engine or potentially gave rise to a GRB
(Berger et al. 2003b).

In the majority of cases we find expansion velocities of $\simlt 0.3c$
(Figure~\ref{fig:snvel}) as compared to $\Gamma\sim 2$ for SN\,1998bw
and $\Gamma\sim 5$ for GRB radio afterglows.  Similarly, the energy
carried by these ejecta can be accounted for in a hydrodynamic
explosion model.  In fact, as seen in Figure~\ref{fig:energy} SNe Ib/c
are significantly less energetic in terms of fast ejecta compared to
GRBs.  Thus, GRBs and the vast majority of SNe Ib/c do not share a
common energy source.  We can therefore rule out models of GRBs or SNe
which suggest a significant overlap (e.g.~Lamb et al. 2003).

\section{Future Directions}
\label{sec:future}

The recent recognition of a new class of cosmic explosion, the X-ray
flashes (XRFs), provides an opportunity to extend the analysis
presented above.  These transients are defined as those with  ${\rm
log}[S_X(2-30\,{\rm keV})/S_\gamma(30-400\,{\rm keV})]>0$, where $S_X$
and $S_\gamma$ are the fluences in the X-ray and $\gamma$-ray bands,
respectively; i.e.~the peak in the $\nu F_\nu$ spectrum lies in the
X-ray band.  With the exception of a significantly lower peak energy,
XRFs share similar properties (e.g.~duration, fluence) with GRBs.

Recent detections of XRF afterglows indicate that they likely arise at
cosmological distances: they exhibit interstellar scintillation effects
similar to those observed in GRB radio afterglows, they are associated
with faint compact galaxies similar to the general population at
$z\sim 1$, and in one case (XRF\,020903), a redshift of 0.25 has been
measured.  

Using the measured redshift, the prompt energy release of XRF\,020903
is only $\sim 2\times 10^{49}$ erg, significantly lower than GRBs.
The difference is even more pronounced when we consider that the
spectrum peaked at $\sim 5$ keV.  Thus, it is possible that
XRF\,020903 (and perhaps all XRFs) produce a negligible amount of
highly relativistic ejecta, maybe as a result of higher baryon
contamination in the ejecta.  However, preliminary results indicate
that the total relativistic output of XRF\,020903, measured from the
radio afterglow in the usual manner, is similar to that of GRBs
(Figure~\ref{fig:energy}; Soderberg et al. 2003).  If so, XRFs may in
fact share a common origin with GRBs.

To assess this possibility it is crucial to obtain a large sample of
XRFs with measured redshifts.  This, along with continued monitoring
of GRBs and their afterglows (especially in the Swift era) and
continued radio observations of SNe Ib/c, will allow us to determine
more accurately the true diversity of cosmic explosions and the
fraction of stellar deaths in each channel.

\bigskip\noindent
{\it Acknowledgments} \\ I would like to thank the conference
organizers, C.~Wheeler, P.~Kumar, and P.~Hoflich, and my many
collaborators, in particular S.~Kulkarni and D.~Frail

\end{document}